\documentclass[11pt]{article}

\RequirePackage{amsthm,amsmath,amsfonts,amssymb, bm, mathtools, graphicx, float, multirow, enumitem, url, bbm}
\RequirePackage[authoryear]{natbib}
\RequirePackage[colorlinks,citecolor=blue,urlcolor=blue]{hyperref}

\usepackage{amsmath}
\usepackage{amsfonts}
\usepackage{bbm}
\usepackage{bm}
\usepackage{graphicx}
\usepackage{enumerate}
\usepackage{url}

\usepackage{caption}
\usepackage{setspace}
\usepackage{wrapfig}
\usepackage{appendix}
\usepackage{float}

\usepackage[export]{adjustbox}
\usepackage{wrapfig}
\usepackage{framed}
\usepackage{xcolor}
\usepackage{authblk}
\begin{document}

\def\spacingset#1{\renewcommand{\baselinestretch}%
{#1}\small\normalsize} \spacingset{1}


  \title{\bf Model-based indicators for co-clustered environments and species communities}
    \author[1]{Braden Scherting}
    \author[2]{Otso Ovaskainen}
    \author[3]{Tomas Roslin}
    \author[1]{David B. Dunson}
    \affil[1]{Department of Statistical Science, Duke University}
    \affil[2]{Department of Biological and Environmental Science, University of Jyv\"askyl\"a; P.O. Box 35, 40014 Jyv\"askyl\"a, Finland.}
    \affil[3]{Department of Ecology, Swedish University of Agricultural Sciences (SLU); Ecosystems and Environment Research Programme, Faculty of Biological and Environmental Sciences, University of Helsinki, Finland }
  \maketitle

\begin{abstract}
Accurate biodiversity monitoring is essential for effective environmental policy, yet current practices often rely on arbitrarily defined ecosystems, communities, and ad‑hoc indicator species, limiting cost‑efficiency and reproducibility. We present a model‑based framework that infers ecological sub‑communities and corresponding indicators in terms of habitat and species from species survey data, such as large‑scale arthropod abundance data used here as example. Environments and species are co-clustered using Bayesian decoupling for Poisson factorization. Latent, hierarchical regression relates observable habitat features to each subcommunity. Additionally, we propose a novel, model-based ranking of indicator species based on the learned subcommunities, generalizing classical approaches. This integrated approach  motivates model-based ecosystem classification and indicator species selection, offering a scalable, reproducible pathway for biodiversity monitoring and informed conservation.
\end{abstract}

\noindent%
\vfill


\section{Introduction: Ecological Communities and Indicators}

To inform environmental policies and guide conservation action, we need accurate information on the status and trends in the environment. This task is massively complex, since nature is made up of so many different environments and so many species. The question is then what units to survey and monitor to obtain a comprehensive and representative view of environmental change \citep{goodsell2025moving}.  

Given wide variation in species communities and environmental conditions, a common approach to surveying the environment 
is to define a set of "types" to focus on. In terms of ecosystems, several countries have committed to assessing the threat levels of regional ecosystems, adhering to the IUCN’s Red List of Ecosystems (RLE; \cite{keith2015iucn}). Similarly, the assessment of the threat status of species (so called ``red listing'' of species) is frequently based on changes in the extent of the ecosystems on which they depend \citep{iucn2003guidelines}.

In terms of species, there have been similar approaches to defining ``types'' of communities. Given the effort required to document the occurrence and abundance of each individual species, the occurrence and status of a given type of community has often been assigned to informative proxies, i.e., one or several ``indicator species'', which through their presence or abundance are assumed to reveal the wider state of many (undocumented) species in the surrounding community \citep{carignan2002selecting}.

What renders this approach problematic is the general lack of objective and quantitative criteria for defining an ``ecosystem'', a ``community'' or ``an indicator species'' in the first place. As a case in point, the area of Canada (ca $10^{6}$ $\text{km}^{2}$) has been divided into 1,027 separate ecological units called ecodistricts \citep{statcan2025}. In terms of species communities, Canada is home to some 80,000 described species, excluding bacteria and viruses \citep{mcgillBiodiversity2025}. With a massive proportion of species that have yet to be discovered, the true number could be twice that \citep{mcgillBiodiversity2025}. 

When faced with this massive variation, class limits for ecosystems and communities are typically defined by expert opinion, with finer classes formed within some subcategories and wider in others (e.g. \cite{envirofi2022}). Indicator species are frequently selected \emph{ad hoc}, without quantitative proof of their indicator value \citep{siddig2016ecologists}. As the usual workflow, ecologists first select environments of interest (e.g. "old growth spruce forests", then select species indicating them (e.g. wood-decaying fungi, such as \emph{Pycnoporellus fulgens} and \emph{Phlebia centrifuga}). In the worst case, such approaches will result in low cost-efficiency: we will spend much effort in documenting the presence of a non-informative indicator species to establish the status of a poorly-defined ecosystem only recognizable by an expert observer -- where documenting the environmental conditions in the first place would have yielded a better description of the state of the environment \citep{bal2018selecting}.   

As an obvious alternative, we should advance from arbitrary definitions towards model-based, objective criteria. Several promising steps have recently been taken in this direction. For environments, methods based on automatic classification of remotely-sensed data can produce repeatable, verifiable ecosystem types \citep{wulder2004high, pettorelli2018satellite} but will obviously need ground proofing. For species communities, quantitative criteria for establishing Regions of Common Profile contribute to objective, repeatable classifications of "ecosystems" \citep{foster2013modelling}. Species archetype models \citep{dunstan2011model, hui2013mix} identify a small number of characteristic, hypothetical species in a mixture modeling framework. Although archetypal species are conceptually related to indicator species, they bear less utility for future monitoring. Similarly, model-based ordination \citep{hui2015model} is useful for grouping samples and species, but falls short of identifying indicators for either. What's more, most of these alternative approaches are underdeveloped for applications involving tens of thousands of species.

Modern technologies for scoring biodiversity now render ecosystem classification, community classification and the selection of indicator species ripe for a reassessment. Large-scale biodiversity data are becoming increasingly available as technology improves \citep{bush2017connecting, hartig2024novel}, with examples such as the Global Spore Sampling Project \citep{ovaskainen2024global}, Insect Biome Atlas \citep{miraldo2025data}, and Lifeplan \citep{hardwick2024lifeplan}. Such data bring the main part of previously unseen organisms into the realm of documented communities. Rather than resorting to conjecture regarding how well "indicator species" represent the full community, we may thus address this empirically. Drawing on our improved knowledge regarding the composition and state of the full community, we may thus use quantitative methods to select indicator species, then ask how well these proxies represent the rest.    

In this paper, we propose a combined modeling and estimation scheme for learning arthropod niche partitions and subcommunities from abundance data on arthropods, as the likely most diverse members of terrestrial species communities \citep{mora2011many, stork2018many}. Drawing on the largest data set of equally-sampled, individually identified arthopod communities available to date (\url{https://biodiversitygenomics.net/projects/gmp/}; \cite{seymour2024global}), we characterize the subcommunities by their members, the habitat and geography, and through sets of representative species (indicators). Rather than first defining environments of interest, then selecting species to indicate them, we do both at the same time: we simultaneously cluster environments and species, then ask what species indicate what environments and communities. By drawing on external data, we also probe for the environmental conditions indicating each environmental cluster. More specifically, we ask:
\begin{itemize}
   \item[1) ] How can Canadian arthropods be partitioned into ecologically-relevant subcommunities? 
   \item[2) ] To what extent does this partition reflect observed environmental gradients? Can subcommunity presence or species occurrence be predicted using environmental indicators?
   \item[3) ] What species indicate each subcommunity? Can subcommunity presence or species occurrence be predicted using indicator species?
\end{itemize}

\section{Materials and Methods}

\subsection{Empirical material}
\begin{figure}
\centering
\includegraphics[width=.5\linewidth]{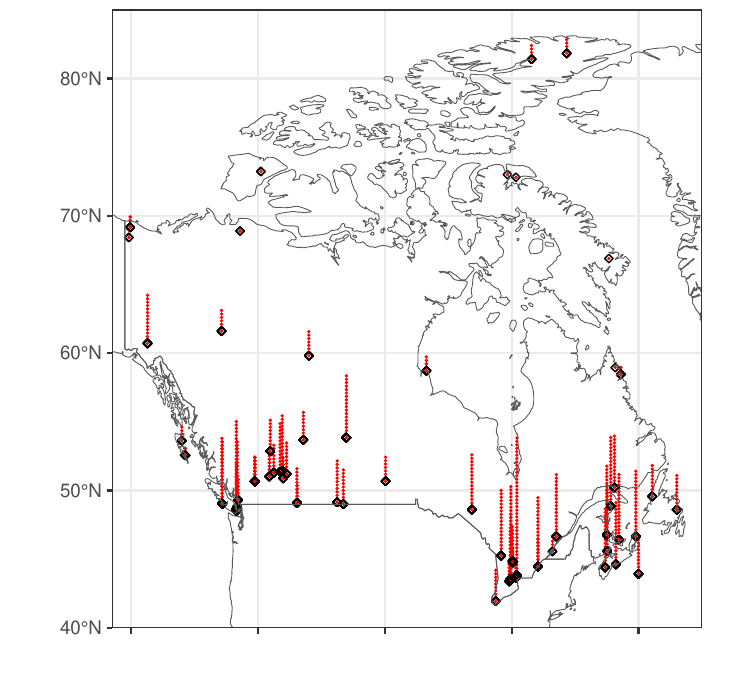}
\caption{Site and sample locations in Canada from the Global Malaise Trap Program (GMTP). Sites, denoted by black diamonds, are distributed throughout Canada, with most situated below $55^\circ$ N. Samples are denoted by red points stacked at corresponding sites.}
\label{fig:explore_map}
\end{figure}

\begin{figure}
\centering
\includegraphics[width=.75\linewidth]{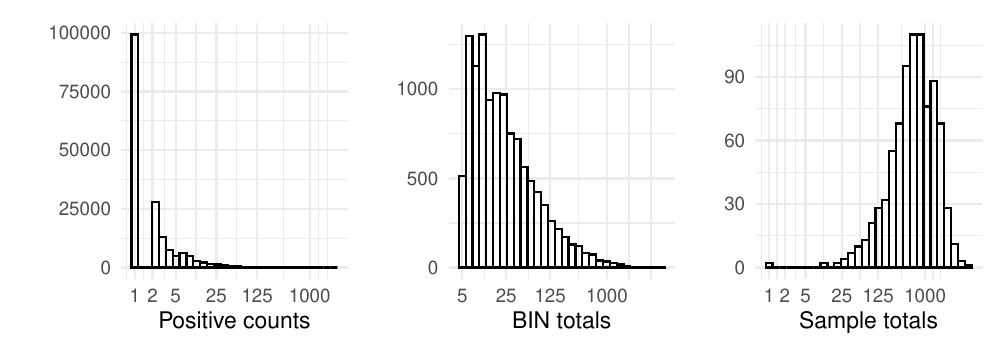}
\caption{Log-scale nonzero counts (left) and empirical marginals (center, right) for the Canada GMTP data, including species present five or more times.  Both small counts (e.g., 1,2,3) and very large counts ($>100$) occur frequently. The distribution of species (BIN) totals has a long right tail---some species are much more common and abundant than others. }
\label{fig:explore_hists}
\end{figure}

We analyze data from the Global Malaise Trap Program (GMTP), a large-scale endeavor aimed at surveying global terrestrial arthropod communities using standardized protocols \cite{seymour2024global}. Malaise traps operate by obstructing the paths of walking or flying arthropods, funneling individuals into an ethanol-filled collection bottle. In this analysis, we refer to each Malaise trap bottle as a sample; in most cases several samples were collected at each site. The GMTP protocol dictated weekly collection events. Specimens collected over the course of each week were cataloged and DNA barcoded by the Centre for Biodiversity Genomics at the University of Guelph. The bioinformatics pipeline employed assigned a Barcode Index Number (BIN) to each specimen, which functions as a proxy for species identification. This classification scheme is critical because approximately 75\% of species captured do not belong to a named species and some 47\% have no named genus. Samples from Canada (n=834; Fig. \ref{fig:explore_map}) alone account for 42232 distinct BINs. Among these, we limit our focus to species that occur in at least five samples (p=11682; Fig. \ref{fig:explore_hists}). The processing pipeline is unique in that each specimen is individually DNA barcoded. Therefore, the data are specimen counts per sample, per BIN. Arranged in a matrix $Y$ with samples along the rows and BINs (henceforth, ``species'') along the columns, the data are $98.2\%$ sparse, and the largest single count is 2602 (unnamed species, genus \textit{Cricotopus}). On average, each species is present in 1.8\% of samples. The simultaneous sparsity and heterogeneity pose a considerable modeling challenge. To supplement these data, we also extract fractional land cover classification data within 100 meters of each site \citep{canlandcover2010}. The land cover types are collapsed into ten categories: barren, taiga, urban, deciduous mix, wetland, shrub, grass, water, cropland, and polar scrub.

\subsection{Statistical methodology}

Our analysis comprises three components: 1) a scalable and robust model for high-dimensional count data styled after hierarchical Poisson-gamma factorization that incorporates covariate information, 2) a sparse estimation procedure for translating latent factors to clusters, and 3) a new, model-based metric for identifying and prioritizing indicator species without a predefined or sharp clustering of samples. 

\subsubsection{Poisson factorization}
We employ hierarchical nonnegative Poisson factorization \citep{cemgil2009bayesian}, a naturally flexible, scalable, and interpretable framework, to model arthropod counts. Flexibility, scalability, and interpretability are direct consequences of the model structure. Arranging specimen counts in a samples-by-species matrix $Y\in \mathbb{N}_0^{n\times p}$, we posit that each count $y_{ij}$ is Poisson and conditionally independent given $k\ll n,p$ latent sample factors $\bm\omega_i$ and species factors $\bm\gamma_j$,
\begin{align*} y_{ij}\sim\text{Poisson}\left(  \bm\omega_i^\top \bm\gamma_j \right).
\end{align*}
The conditional Poisson specification is remarkably flexible, provided that appropriate priors are placed on $\bm\omega_i$ and $\bm\gamma_j$. The rate is given by the sum of nonnegative latent factors $\sum_{l=1}^k \omega_{il}\gamma_{jl}$, so if either $\omega_{il}$ or $\gamma_{jl}$ is small, the contribution of the $l$th factor is curtailed.

Gamma priors are effective at promoting latent factors values close to zero while admitting occasional large values; specific additional model components for zero-inflation or overdispersion are not necessary. Gamma priors also underpin efficient posterior computation, which relies on an alternative model specification:
\begin{align}
    y_{ij}=\sum_{l=1}^k y_{ijl}, & && y_{ijl}\sim\text{Poisson}\left( \omega_{il}\gamma_{jl} \right),
    \label{eq:factorSpecificCounts}
\end{align}
where $y_{ij1},\dots,y_{ijk}$ are latent, conditionally independent factor-specific counts. These latent counts partition each observed count, attributing each portion to a specific factor. Given $y_{ij}$, the factor-specific counts are multinomial-distributed with rates $\left\{\omega_{il}\gamma_{jl} / \bm\omega_i^\top \bm\gamma_j\right\}_{l=1}^k$. The conditional rate of $y_{ijl}$ is a product of scalars, so Poisson-gamma conjugacy applies. Furthermore, because all factor-specific counts are zero when $y_{ij}=0$, the imputation need only be performed over $\{(i,j): y_{ij}>0\}$. Hence, the number of multinomial imputations needed for posterior computation or parameter estimation (typically the most expensive operation) scales with the number of nonzero counts, and the full matrix $Y$ need not be held in memory, a well-recognized property of Poisson-gamma factorization \citep{gopalan2015scalable}. This property is necessary when modeling very large community data.

Latent factors are restricted to be positive and combine additively, so each count is explained by aggregating over the various factors. Being positive, the factors cannot cancel one another out, which further promotes sparsity---additional active factors only increase the expected count. Hence, the partition induced in equation \ref{eq:factorSpecificCounts} is usually sparse, a property we rely on to define subcommunities. Furthermore, in contrast to latent Gaussian factor models, this factorization is only permutation- and scale invariant but does not favor challenging rotational invariance \citep{wang2012nonnegative, donoho2003does}. Scale invariance can be resolved by using priors with constrained support, and permutations can be detected from posterior samples. Poisson factorization and variations thereof have been used in diverse application areas, including content recommendation \citep{gopalan2015scalable, zhou2012beta}, mutational signature analysis \citep{rosales2017signer, zito2024compressive}, and community ecology \citep{scherting2024inferring}.

\subsubsection{Sparse estimation}

In this context, species-specific factors $\bm\gamma_j$ are usually approximately sparse, meaning a subset of the elements contribute negligibly to its norm. The approximate sparsity in $\bm{\gamma}_j$ signals a species' niche preferences and can therefore be used to relate it to other species with similar preferences. However, because the sparsity is approximate, identifying species groups on the basis of preference in a consistent and tractable fashion is extremely challenging, especially when when marginal abundance varies widely across species. Thresholding $\Gamma$ by a fixed value will fail to identify preferences of rare species without discriminating between preferences of common species. 

To summarize species preferences, we propose estimating patterns of exact sparsity in the species factor matrix using Bayesian decoupling (BD). Bayesian decoupling divorces data modeling from sparse estimation by first fitting models that explain data and reflect prior beliefs without sparse prior. In a second stage, a loss that induces exact sparsity is crafted and minimized to obtain sparse estimates of parameters \citep{hahn2015decoupling, li2025bayesian}. BD owes much of its conceptual appeal and computational convenience to discarding priors that promote exact sparsity. 

\cite{bolfarine2024decoupling} consider BD for Gaussian factor analysis but seek sparsity in the induced covariance matrix. We define a loss that promotes sparsity in the species factor loadings matrix directly. Under the Poisson-gamma factorization model, the marginal posterior on species and sample factors is
\begin{align*}
    p(\Omega, \Gamma\mid Y)\propto p(Y\mid \Omega, \Gamma)p(\Omega)p(\Gamma),
\end{align*}
where $p(Y\mid \Omega, \Gamma) = \prod_{ij} \text{Poi}(y_{ij}; \omega_i^\top \gamma_j)$. The decoupled, sparse estimate of species loadings $\hat{G}$ minimizes the posterior expectation of
\begin{align*}
    \mathcal{L}_\lambda \left(\Omega, \Gamma, G\right) = D(\Omega\Gamma^\top, \Omega G^\top) + \lambda\mathcal{P}(\Gamma, G)
\end{align*}
subject to $g_{jl}\geq 0$. The penalty $\mathcal{P}(G, \Gamma)$ promotes sparsity while the divergence $D(\Omega\Gamma^\top, \Omega G^\top)$ penalizes deviations between the original and sparse factorizations. In particular, we choose
\begin{align*}
    D(\Omega\Gamma^\top, \Omega G^\top) = ||\Omega\Gamma^\top - \Omega G^\top ||^2_F,
\end{align*}
and 
\begin{align*}
    \mathcal{P}(G, \Gamma) = \sum_{j,l}\frac{\bar{\gamma_j}}{\gamma_{jl}} g_{jl}
\end{align*}
The posterior expected loss is
\begin{align*}
    \mathbb{E}\left[ \mathcal{L}_\lambda \left(\Omega, \Gamma, G\right)\mid Y\right]&= \mathbb{E} \left[ \text{tr}\left\{ \left(\Omega\Gamma^\top  - \Omega G^\top\right)^\top \left(\Omega\Gamma^\top  - \Omega G^\top\right) \right\}\Big|\ Y\right] + \lambda\mathbb{E}\bigg[\sum_{j,l}  \frac{\bar{\gamma_j}}{\gamma_{jl}}g_{jl}\bigg|\ Y\bigg] \\
    &=\text{tr}\left\{ G\ \mathbb{E}\left[\Omega^\top\Omega \mid Y\right]G^\top - GB^\top - BG^\top \right\} + \lambda\sum_{j,l} \mathbb{E}\left[ \frac{\bar{\gamma_j}}{\gamma_{jl}} \bigg|\ Y\right]g_{jl} + \text{cst}_G, 
\end{align*}
where $B= \mathbb{E}\left[ \Gamma\Omega^\top \Omega \mid Y\right]$ and $\text{cst}_G$ is a term constant in $G$. Let $A$ be the principal square root of $\mathbb{E}\left[ \Omega^\top \Omega \right]$. Then,
\begin{align*}
     \mathbb{E}\left[ \mathcal{L}_\lambda \left(\Omega, \Gamma, G\right)\mid Y\right]&= ||A^{\dagger}B^\top - AG^\top||_F^2 + \lambda\sum_{j,l} \mathbb{E}\left[ \frac{\bar{\gamma_j}}{\gamma_{jl}}\bigg|\ Y\ \right]g_{jl} + \text{cst}_G.
\end{align*}
We define 
\begin{align*}
    \mathcal{L}_\lambda (G) = ||A^{\dagger}B^\top - AG^\top||_F^2 + \lambda\sum_{j,l} \mathbb{E}\left[ \frac{\bar{\gamma_j}}{\gamma_{jl}}\right]g_{jl} 
\end{align*}
and take $\hat{G} = \underset{G\geq 0}{\text{argmin}} \ \mathcal{L}_\lambda (G)$ as the sparse estimate. 

\subsubsection*{Including $\Omega$} An alternative loss can be formulated around $||\Gamma - G||$ or, generically, $D(\Gamma,G)$ directly rather than $||\Omega\Gamma^\top - \Omega G^\top||$. Including $\Omega$ in this way has two key advantages: 1) penalizing $\Omega\Gamma^\top - \Omega G^\top$ reflects the modeling premise that $\Omega G^\top$ should match $Y$ and 2) correlation among factor dimensions enters the loss. We discuss each in turn.

Consider the loss incurred by approximating a future observable $\tilde Y\sim\text{Pois}(\Omega\Gamma^\top)$ by $\Omega G^\top$,
    \begin{align*}
        \mathcal{L}(\tilde Y, \Omega, \Gamma, G) &= || \tilde{Y}-\Omega G^\top ||_F^2 + \lambda\mathcal{P}(\Gamma, G).
    \end{align*}
    In expectation,
    \begin{align*}
        \mathbb{E}_{\tilde Y}\left[\left.\mathcal{L}_\lambda(\tilde Y,\Omega, \Gamma, G)\right| \Omega, \Gamma\right] 
        &= || \Omega\Gamma^\top-\Omega G^\top ||_F^2 + \lambda\mathcal{P}(\Gamma, G)  + \text{cst}_G.
    \end{align*}
    Hence, the chosen loss arises as the expected loss under the relevant predictive distribution, up to a term constant in $G$, and is therefore faithful to the modeling objective of approximating $Y$. 

    Imposing nonnegativity tends to induce strong negative correlations between factors, which are relevant to consider when estimating $\Gamma$. Absent $\Omega$, the expected loss takes an overly simple form, 
    \begin{align*}
        \mathbb{E}\left[||\Gamma - G||^2_F\mid Y \right]&\propto_G ||\mathbb{E}[\Gamma\mid Y] - G||^2_F \\
        &= \sum_{j,l} |\bar\gamma_{jl} - g_{jl}|^2
    \end{align*}
    which fails to account for this correlation. By contrast, notice that $A = 
    \Sigma_\Omega+ \mathbb{E}\left[\Omega\right]^\top\mathbb{E}\left[\Omega\right]$, where $\left[\Sigma_\Omega\right]_{ll'}=\text{cov}(\bm{\omega}_l, \bm{\omega}_{l'})$, enters our proposed loss directly.
\subsubsection*{Norm choice} An alternative loss specification is 
\begin{align*}
    D_{Pois}(\Omega\Gamma^\top, \Omega G^\top) = -\sum_{ij} \bm{\omega}_i^\top \bm{\gamma}_j \log\bm{\omega}_i^\top \bm{g}_j + \bm{\omega}_i^\top \bm{g}_j.
\end{align*}
This loss is proportional in $G$ to the Poisson log-likelihood function with ``data'' $\Omega\Gamma^\top$ and rate $\Omega G^\top$. Choosing such a model-aligned loss is appealing, but poses conceptual and computational challenges in this case. This loss cannot be related to the expectation of $D_{Pois}(\tilde Y, \Omega G^\top)$ in closed form for general $\Omega G^\top$. Furthermore, although Poisson-loss nonnegative matrix factorization has been applied broadly, penalized Poisson NMF remains a challenging optimization problem. Therefore, the Frobenius loss is preferred.

\subsubsection*{Penalty choice} The $l_1$ norm serves to induce exact sparsity while remaining computationally tractable. However, applying shrinkage uniformly risks downward biasing species factors that should be nonzero while totally shrinking signals corresponding to rare species. To mitigate these risks, we adopt a penalty belonging to the class of reweighted $l_1$ penalties described by \cite{li2025bayesian} which takes the form
\begin{align*}
    \mathcal{P}(G, \Gamma) &= \sum_{j,l}w_{il}\left(\Gamma\right) g_{jl}\\
    w_{il}\left(\Gamma\right)&=\frac{\bar{\gamma}_j}{\gamma_{jl}}.
\end{align*}
This choice prioritizes within-species sparsity of the factors (i.e., each $\bm{g}_j$ has $k'<k$ nonzero entries) rather than global sparsity in $G$ by assigning weights inversely proportional to the estimated signal strength $\gamma_{jl}^{-1}$, scaled by species' average signal strength $\bar{\gamma}_j$. The species-specific rescaling ensures rare species are penalized similarly to common species. Posterior expectations of the weights are straightforward to compute, and optimization remains straightforward. 

The tuning parameter $\lambda$ must also be chosen to balance sparsity against hypothetical predictive accuracy. To do so, we employ an additional, conceptual constraint on $G$: every species must load on at least one factor, or $\mathbf{1}^\top \bm{g}_j>0$ for all $j$. The posterior benchmarking criterion of \cite{li2025bayesian} can be employed simultaneously, though we find the former requirement to be much stricter. Thus, $\lambda$ is chosen as large as possible such that no rows of $G$ are totally sparse.  

\subsubsection*{Optimization considerations} The optimization problem is solved using blockwise coordinate descent \citep{kim2014algorithms}. Let $Z = A^{-1}B^\top$ and reexpress the expected loss in terms of columns of $G$ as 
\begin{align*}
    \mathcal{L}_\lambda (G) &= \Big|\Big|Z^{(l)} - \textbf{a}_l \bm{g}_l^\top \Big|\Big|^2_F + \lambda \sum_{j=1}^p w_{jl}g_{jl}+ \lambda \sum_{k'\neq k}\sum_{j=1}^p w_{jl'}g_{jl'},
\end{align*}
where $\bm{a}_l$ and $\bm{g}_l$ are the $l$th columns of $A$ and $G$, and $Z^{(l)} = Z - \sum_{l'\neq l} \bm{a}_{l'} \bm{g}_{l'}^\top$. We then iteratively solve the $l=1,\dots,k$ subproblems 
\begin{align*}
    \bm{g}_l&\leftarrow \underset{G\geq 0}{\text{argmin}} \ \Big|\Big|Z^{(l)} - \textbf{a}_l \bm{g}_l^\top \Big|\Big|^2_F + \lambda \sum_{j=1}^p w_{jl}g_{jl}
\end{align*}
Letting $\mathcal{L}^{(k)}\left(\bm{g}\right) = \Big|\Big|Z^{(k)} - \bm{a}_k \bm{g}^\top \Big|\Big|^2_F + \lambda \sum_{j=1}^p w_{jl}g_{jl}$, we have
\begin{align*}
    \frac{\partial \mathcal{L}^{(l)}\left(\bm{g}\right)}{\partial g_j} &= 2\left( g_j \bm{a}_l^\top \bm{a}_l - Z^{(l)\top}_j \bm{a}_l \right) + \lambda w_{jl}
\end{align*}
Therefore, subproblem solutions are given by
\begin{align*}
    g_{jl}\leftarrow  \frac{\text{max} \left\{Z^{(l)\top}_j \textbf{a}_l-\frac{1}{2}\lambda w_{jl}, 0\right\} }{\textbf{a}_l^\top \textbf{a}_l}
\end{align*}

\subsubsection{Introducing habitat information}
Because factors are nonnegative and combine additively to determine the mean, the different dimensions can be interpreted as distinct, composite environmental drivers. Sample factor scores $\bm{\omega}_i$ indicate which of the features are present in a given sample and to what extent, and species loadings indicate preferences toward each factor. When the specific environmental drivers cannot be identified using covariate information, the factors and loadings are nonetheless valuable for coherently associating different species and sites. 

Let $y_{i\cdot l}=\sum_j y_{ijl}$ and $y_{\cdot\cdot l} = \sum_i y_{i\cdot l}$. Motivated by the fact that 
\begin{align*}
    \left(y_{1 j l},\dots, y_{nj l}\mid \text{---}\right)\sim \text{Multinomial}\left( y_{\cdot j l}, \bm{\omega}_l\right)\implies    \left(y_{1\cdot l},\dots, y_{n\cdot l}\mid \text{---}\right)\sim \text{Multinomial}\left( y_{\cdot\cdot l}, \bm{\omega}_l\right),
\end{align*}
we consider a logistic-normal model for sample factors $\bm\omega_{l}$ by letting
\begin{align*}
    \omega_{il} &= \frac{e^{\eta_{il}}}{\sum_i e^{\eta_{il}}}\\
    \eta_{il}&\sim \text{Normal}\left(\mu_{il}, \tau^{-2}\right)
\end{align*}
We model $\mu_{il}$ as $x_i^\top \beta_l$ when covariates are available and fix $\mu_{il}=0$ otherwise. The latent Gaussian precision $\tau^2$ is a tuning parameter that controls the smoothness of factors across samples, similar to the Dirichlet concentration parameter. Constraining $\sum_i \omega_{il}=1$ resolves scale ambiguity between $\bm\omega_l$ and $\bm\gamma_l$. 

Although this model specification uses many of the tools developed for multinomial logistic regression, it differs in a few key ways: 1) samples are the unit of normalization or multinomial ``categories'', 2) both the category-specific counts $y_{i\cdot l}$ and totals $y_{\cdot \cdot l}$ are latent, 3) $\eta_{il}$ is permitted to deviate from $\mu_{il}$, and 4) $x$ varies across categories rather than $\beta$. The transformation (softmax) is invariant under constant shifts to $\eta$ across all $i$. This invariance can be resolved in a number of ways; we remove the intercept and employ informative priors on regression coefficients.

Following \cite{held2006bayesian}, the conditional likelihood of $\eta_{il}$ given $\eta_{-il}$ and $Y$ is
\begin{align*}
    \ell \left(\eta_{il}\mid \left\{\eta_{-il}\right\}, Y\right) &= \prod_{j=1}^p \left[\frac{\exp\left(\psi_{il}\right)}{1 + \exp\left(\psi_{il}\right)}\right]^{y_{ijl}}\left[\frac{1}{1 + \exp\left(\psi_{il}\right)}\right]^{y_{\cdot j l}} \\
    &= \left[\frac{\exp\left(\psi_{il}\right)}{1 + \exp\left(\psi_{il}\right)}\right]^{y_{i\cdot l}}\left[\frac{1}{1 + \exp\left(\psi_{il}\right)}\right]^{y_{\cdot \cdot l}},
\end{align*}
which admits an augmented form with Poly\'{a}-Gamma auxiliary variables \citep{polson2013bayesian}.

\subsubsection{Model-based indicator species}
We frame indicator species selection as an estimation problem. The general task is to select a subset of species represented by their indices $\mathcal{J}\subset \{1,2,\dots,p\}$ such that reduced data $Y_{\mathcal{J}}=[\bm{y}_{j_{r}}]$ for $r\in \mathcal{J}$ carry salient information about the unreduced data $Y$. In general, the manner of choosing $\mathcal{J}$ is highly context dependent. 

A canonical measure of a species' quality as an indicator is the ``indicator value'' (IndVal) \citep{dufrene1997species}. To apply IndVal, samples are first each assigned to distinct clusters, which are either defined in advance based on management criteria or are estimated in a preprocessing step. An indicator species will be assigned to each cluster based on the corresponding IndVal score. Let $r_i\in\{1,\dots,K\}$ denote the cluster label of sample $i$ and $n_l$ the number of samples assigned to cluster $l$. In the original IndVal formulation, clusters are mutually exclusive, a requirement we will relax. In this notation, the IndVal score for species $j$ in cluster $r$ is the product of a ``concentration'' score $A_{jl}$ and a ``fidelity'' score $B_{jl}$, where 
\begin{align*}
    A_{jl} &= \frac{n_l^{-1}\sum_{\{i:r_i=l\}} y_{ij}}{\sum_{l'=1}^K n_{l'}^{-1}\sum_{\{i:r_i=l'\}} y_{ij}} \\ 
    \\
    B_{jl} &= \frac{\sum_{i=1}^n \mathbbm{1}(r_{i}=l) \mathbbm{1}(y_{ij}>0)}{n_l},
\end{align*}
where $y_{ij}$ is the abundance of species $j$ in sample $i$. The concentration term $A$ encodes the notion that a good indicator species for sample type $l$ should have high abundance in $l$-type samples relative to samples of other types. The fidelity term says that a good indicator should be present in most samples of type $l$. The terms are at most $1$ when $j$ is present in only and all $l$-type samples. 

In many studies, well-defined clusters are not readily available, and inferring distinct clusters in a first stage is both challenging and risks double use of data. Nonetheless, IndVal provides an elegant framework by which to define a decision rule for selecting indicator species based on a joint model of abundance. In the NMF setting, we treat factor dimensions as clusters and factor scores $\bm{\omega}_i=(\omega_{i1},\dots,\omega_{ik})^\top$ as soft clustering assignments. The loss incurred by using species $j$ as the indicator for factor/cluster $l$ is 
\begin{align*}
    \mathcal{L}^{Ind}_{jl} = -\frac{\gamma_{jl}}{\gamma_{j\cdot}} \sum_i \omega_{il}\left[1 - \exp(-\bm{\omega}_i^\top \bm{\gamma}_j)\right],
\end{align*}
where $\gamma_{j\cdot}=\mathbf{1}^\top \bm{\gamma}_j$. To see the correspondence between this loss and the canonical IndVal, consider once again a future observable $\tilde{Y}\sim\text{Pois}\left(\Omega\Gamma^\top\right)$. The model for $\tilde{Y}$ can be equivalently expressed using factor-specific abundances:
\begin{align*}
    \tilde{y}_{ij} &= \sum_{l}\tilde{y}_{ijl}, && \tilde{y}_{ijl}\sim \text{Pois}(\omega_{il}\gamma_{jl}).
\end{align*}
Hence, in expectation with respect to the predictive distribution, the average abundance of $j$ attributable to cluster $l$ is 
\begin{align*}
    \mathbb{E}_{\tilde{y}}\left[ n^{-1}\sum_{i=1}^n \tilde{y}_{ijl} \right] &=\frac{\gamma_{jl}}{n}.
\end{align*}
which leads to the model based definition of $A$, 
\begin{align*}
    \tilde{A} \coloneq \frac{\gamma_{jl}/n}{\sum_{l'=1}^k\gamma_{jl'}/n} = \frac{\gamma_{jl}}{\gamma_{j\cdot}}.
\end{align*}
This definition preserves both the qualitative interpretation of $\tilde{A}$ as a concentration score---a large value of $\gamma_{jl}$ relative to $\gamma_{j\cdot}$ indicates a high expected abundance in $l$-type samples relative to other samples---and the bounding to $[0,1]$, with the maximum $\tilde{A}=1$ indicating that $j$ occurs only in $l$-type samples.

A model-based fidelity score can be obtained similarly by considering the soft clustering assignments $\omega_{il}$ and the probability of presence, 
\begin{align*}
     \tilde{B}& \coloneq\mathbb{E}_{\tilde{y}} \frac{ \sum_{i=1}^n \omega_{il} \mathbbm{1}(\tilde{y}_{ij}>0)}{\sum_{i=1}^n \omega_{il}} \\
     &= \sum_{i=1}^n \omega_{il} \left[1 - \exp(-\bm{\omega}_i^\top \bm{\gamma}_j)\right], 
\end{align*}
which is also contained in $[0,1]$. The maximum is achieved when the probability of occurrence is exactly 1 for all samples $i$ such that $\omega_{il}>0$. The product $\tilde{A}_{jl}\tilde{B}_{jl}$ forms the model-based IndVal score (MB-IndVal). The negative score is our chosen loss, $\mathcal{L}^{ind}_{jl}=-\tilde{A}_{jl}\tilde{B}_{jl}$. For each factor $l=1,\dots,k$, the optimal indicator is 
\begin{align*}
    j^*_l = \underset{j\in\{1,\dots, p\}}{\text{argmin}} \mathbbm{E}\left[\mathcal{L}^{ind}_{jl}\mid Y\right],
\end{align*}
which can be found by simply searching through candidate species. In settings where $m$ indicators per factor are desired, it is natural to order species by $\mathcal{L}^{ind}_{jl}$ and select the top $m$, or select those that are easiest to collect and identify in the field. 

This approach is particularly well-suited to settings where clusters are not obvious or known in advance. Without additional information or modeling effort put towards interpreting the learned clusters in the context of known environmental features, using the estimated indicators for monitoring specific ecosystems is challenging. However, indicator species identified in this manner provide two important functions in large-scale semi-autonomous biomonitoring settings: 1) aiding in the ecological description of undescribed or lesser-known species through well-known focal species, and 2) selecting an easy-to-identify, abundant subset of species that reflect dominant environmental trends for lower-cost future sampling. This is particularly relevant in the context of GMTP data, because many of even the most common species are known only from DNA.

\subsubsection{Hierarchical priors and computation strategies}
The sparse estimation procedure, indicator species scoring, and model for covariate effects are agnostic to the choice of prior for $\Gamma$. We adopt the following:
\begin{align*}
    \gamma_{jl} &\sim \text{gamma}\left(\xi_l, \text{scale}=\theta_l\right) \\
    \xi_{l} &\sim \text{gamma}\left(a_0, \text{scale}=b_0\right) \\
    \theta_l&\sim \text{inverse-gamma}\left(c_0, d_0\right),
\end{align*}
where $a_0=2$, $b_0=1$, $c_0=2$, and $d_0=1$. Simultaneously estimating $\xi_l$ and $\theta_l$ adds flexibility, and data-augmented complete conditionals are available for both \citep{zhou2013negative}. 

This model admits an efficient Gibbs sampler, which we initialize at the approximate maximum a posteriori (MAP) estimate. To initialize, we solve 
\begin{align*}
    \underset{\Gamma, \Omega, \bm\xi, \bm\theta}{\text{argmin}}-\log\left\{p(Y\mid\Gamma, \Omega)p(\Gamma\mid \bm\xi, \bm\theta)p(\bm\xi, \bm\theta)\right\},
\end{align*} 
dropping the prior on $\Omega$ for simplicity and because inference on $\Gamma$ is more sensitive to initialization. The solution is found by iteratively updating each parameter. Closed-form updates for $\Omega$, $\Gamma$, and $\bm\theta$ are available: 
\begin{align*}
    \Omega_{:l} &\leftarrow \Omega_{:l}\ \odot\frac{\left(Y\oslash\hat{Y}\right) \Gamma_{:l}}{\mathbf{1}_p^\top \Gamma_{:l}}  \\
    \Gamma_{:l}&\leftarrow \Gamma_{:l}\ \odot \frac{\left(Y\oslash\hat{Y}\right)^\top \Omega_{:l} + \xi_l - 1}{\mathbf{1}_n^\top \Omega_{:l} + \theta_l^{-1}} \\
    \theta_l&\leftarrow \frac{\mathbf{1}_p^\top \Gamma_{:l} + d_0}{p\xi_l + c_0 + 1}, 
\end{align*}
where $\hat{Y} = \Omega\Gamma^\top$, and $\odot$ and $\oslash$ are elementwise addition and division, respectively. Columnwise normalization of $\Omega$ is enforced at each iteration, i.e., $\Omega_{:l} \leftarrow \Omega_{:l} / \mathbf{1}_n^\top \Omega_{:l} $. Because the gradient of the log posterior with respect to $\xi_l$ involves the digamma function $\psi(x)$, the factor-specific shape parameter $\xi_l$ is updated using a small number of damped Newton steps,

\begin{align*}
    \xi_l \leftarrow \xi_l - \tau \frac{p \left[\psi(\xi_l) + \log \theta_l\right] -\sum_j \log \gamma_{jl} - (a_0-1)/\xi_l + \theta_l^{-1}}{p\psi_1(\xi_l) + (a_0-1)/\xi_l^2},
\end{align*}
where the step size $\tau$ is tuned to ensure $\xi_l>0$. This procedure can be performed rapidly and in parallel to mitigate (but not eliminate) the risk of initializing near an unfavorable mode.

\section{Results}\
To fit the model, we compute 50 approximations to the MAP in parallel and initialize at the one that achieves the highest log posterior. We draw 35,000 samples from the posterior, discarding the first 25,000 and thinning the remaining 10,000 by 10 to respect memory limitations. For most analyses presented here, the factorization rank is fixed at $k=5$. We also consider $k\in\{3,10,15\}$; among these, $k=10$ achieves the lowest WAIC, and $k=5$ is preferred to both $k=3$ and $k=15$. Posterior predictive checks indicate that both sample- and species-specific marginals are well-represented by the model with $k=5$ [Supplement]. Hence, we choose to present results for $k=5$ instead of $k=10$ due to the better interpretability. In this section, we present clusters representing arthropod subcommunities, regions of common profile, and indicators for each cluster based on habitat characteristics and representative species. 

\subsection{Arthropod subcommunities}
\begin{figure}[h]
\includegraphics[width=1\linewidth]{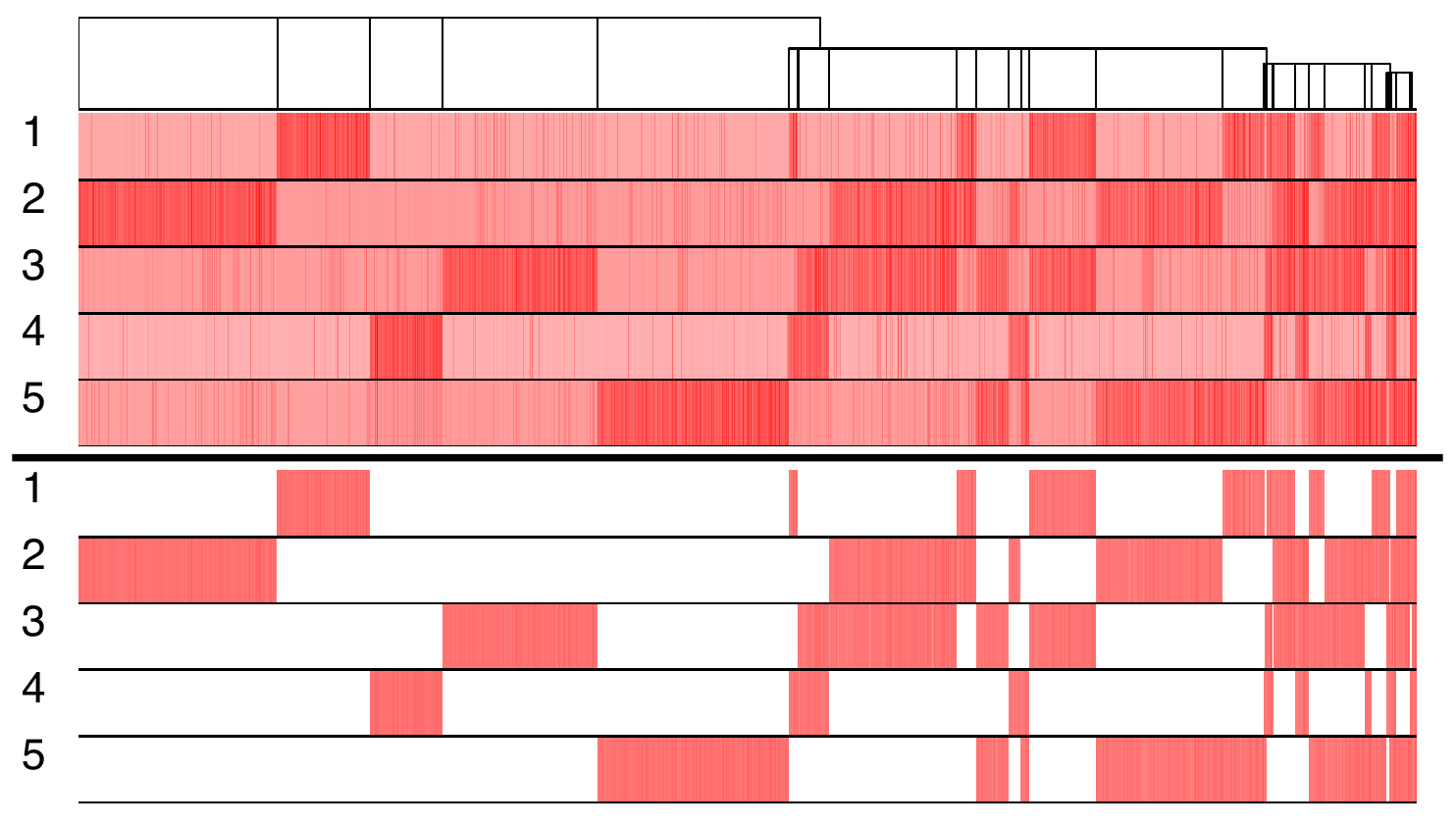}
\caption{Species loadings (k=5) in our analysis of the Canada GMTP data. The top panel represents the posterior means of the loadings, while the bottom panel shows the sparsified decoupling estimator. The columns (species) are arranged in terms of the inferred subcommunities.}
\label{fig:stacked_Gclust}
\end{figure}

Figure \ref{fig:stacked_Gclust} presents the posterior mean of species factors $\Gamma^\top$ and the sparse estimate $\hat{G}^\top$ clustered by sparsity pattern and ordered by the number of nonzero entries in $\hat{g}_j$. Each collection of species that share a sparsity pattern form a subcommunity characterized by preferential co-occurrence. The subcommunities are not mutually exclusive in general, because the form of sparsity promoted by our loss is species- \textit{and} factor-specific, rather than only species-specific (e.g., requiring only one nonzero factor per species). This enables us to identify both 1) collections of species that can be strictly partitioned into non-overlapping subcommunities and 2) collections that cannot. 

Approximately half ($6172/11628$) of species load on a single factor. These species can be effectively clustered in the following sense: if $\mathcal{J}_l$ is the collection of species that load only on factor $l$, then $\bm\omega_l\hat{\bm{g}}_{\mathcal{J}_ll}^\top$ (or $\bm\omega_l\bm\gamma_{\mathcal{J}_ll}^\top$) is a good, rank-one approximation to the counts of species $j\in\mathcal{J}_l$, $$Y_{\mathcal{J}_l}\approx\hat{\bm{g}}_{\mathcal{J}_ll}\bm\omega_l^\top.$$
These subcommunities appear on the left half of Figure \ref{fig:stacked_Gclust}. 

Because species can load on more than one factor, other subcommunities overlap. However, sparsity still persists: 89\% load on two or fewer and 98\% on three or fewer. Species that load on all factors are interesting in that those $j$ for which $\hat g_j$ is nowhere sparse are approximately as likely to appear in one sample as any other. These species are cosmopolitan with respect to the learned subcommunities and corresponding niche partitions, as they can exist in \textit{any} subcommunity habitat type. This perspective on cosmopolitanism is more closely connected to ecosystem functioning than is geographic cosmopolitanism. 

$\hat{G}$ identifies 39 cosmopolitan species (Table \ref{tab:cosmos}). While many of these species are also geographically cosmopolitan---11 of 39 are among the top 1\% most common species, including the Utah funnelweb spider (\textit{Agelenopsis utahana}) and double-striped scoparia moth (\textit{Scoparia biplagialis})--several are comparatively rare. \textit{Syrphus vitripennis}, a species of hoverfly, appears in less than 5\% of samples (38 detections; 55 specimens). Yet indeed, it has been described across the Northern Hemisphere, and hoverflies are generally very widely distributed and migratory \citep{reynolds2024comprehensive}. Another unnamed, much more common Syrphidae (\textit{Melanostoma}, BOLD:AAB2866), is also cosmopolitan, but the other 93 Syrphidae species belong to more specific subcommunities.

\begin{table}[h]
\centering
\caption{Cosmopolitans}
\footnotesize\begin{tabular}{lllll}
\hline
\textbf{BIN} & \textbf{Class} & \textbf{Order} & \textbf{Family} & \textbf{Genus species} \\
\hline
BOLD:ABU5525 & Insecta & Diptera & Chironomidae & \textit{ Limnophyes sp. 14ES }\\
BOLD:AEU6731 & Insecta & Hymenoptera & Braconidae & \textit{ Dinotrema }\\
BOLD:AAI8935 & Insecta & Coleoptera & Latridiidae & \textit{ Cortinicara gibbosa }\\
BOLD:AAB2866 & Insecta & Diptera & Syrphidae & \textit{ Melanostoma }\\
BOLD:AAH3228 & Insecta & Psocodea & Caeciliusidae & \textit{ Valenzuela flavidus }\\
BOLD:AAB5577 & Insecta & Diptera & Syrphidae & \textit{ Syrphus vitripennis }\\
BOLD:ACK3142 & Insecta & Diptera & Cecidomyiidae & \textit{ None }\\
BOLD:ACS7008 & Insecta & Diptera & Phoridae & \textit{ Megaselia arcticae }\\
BOLD:ACM1917 & Insecta & Hymenoptera & Scelionidae & \textit{ Telenomus autumnalis }\\
BOLD:AAP3767 & Insecta & Diptera & Cecidomyiidae & \textit{ None }\\
BOLD:AEW1133 & Insecta & Diptera & Sciaridae & \textit{ Scatopsciara atomaria }\\
BOLD:AAA6280 & Insecta & Hymenoptera & Ichneumonidae & \textit{ Gelis }\\
BOLD:AAV5088 & Insecta & Diptera & Ceratopogonidae & \textit{ Forcipomyia }\\
BOLD:AAC2498 & Insecta & Diptera & Muscidae & \textit{ Helina }\\
BOLD:AAA6213 & Insecta & Hemiptera & Aphididae & \textit{ Macrosiphum }\\
BOLD:AAP2528 & Insecta & Diptera & Keroplatidae & \textit{ Orfelia nemoralis }\\
BOLD:AAG1488 & Insecta & Hymenoptera & Mymaridae & \textit{ Lymaenon }\\
BOLD:AAC0706 & Insecta & Diptera & Chironomidae & \textit{ Dicrotendipes tritomus }\\
BOLD:AAG0891 & Insecta & Neuroptera & Hemerobiidae & \textit{ Hemerobius }\\
BOLD:AAA1518 & Insecta & Lepidoptera & Crambidae & \textit{ Scoparia biplagialis }\\
BOLD:AAC8842 & Insecta & Diptera & Chironomidae & \textit{ Paratanytarsus laccophilus }\\
BOLD:AAP7843 & Insecta & Coleoptera & Cantharidae & \textit{ Malthodes pumilus }\\
BOLD:AAB0090 & Arachnida & Araneae & Agelenidae & \textit{ Agelenopsis utahana }\\
BOLD:AAG6519 & Insecta & Diptera & Ceratopogonidae & \textit{ Atrichopogon }\\
BOLD:AAV5609 & Insecta & Diptera & Cecidomyiidae & \textit{ None }\\
BOLD:AAG3625 & Insecta & Diptera & Cecidomyiidae & \textit{ None }\\
BOLD:AER3363 & Insecta & Diptera & Chironomidae & \textit{ Ablabesmyia americana }\\
BOLD:AAB8787 & Insecta & Diptera & Lauxaniidae & \textit{ Minettia lupulina }\\
BOLD:AAG1704 & Insecta & Diptera & Muscidae & \textit{ Lispocephala erythrocera }\\
BOLD:ACZ5374 & Insecta & Diptera & Anthomyiidae & \textit{ Lasiomma }\\
BOLD:AEV9539 & Insecta & Diptera & Chironomidae & \textit{ Limnophyes asquamatus }\\
BOLD:ABU5545 & Insecta & Diptera & Mycetophilidae & \textit{ Cordyla }\\
BOLD:AAB0079 & Insecta & Diptera & Chironomidae & \textit{ Corynoneura arctica }\\
BOLD:ACX4619 & Insecta & Coleoptera & Scirtidae & \textit{ Contacyphon variabilis }\\
BOLD:AAG2464 & Insecta & Diptera & Anthomyiidae & \textit{ Pegomya }\\
BOLD:ACX5107 & Insecta & Diptera & Cecidomyiidae & \textit{ None }\\
BOLD:AAE4568 & Insecta & Diptera & Chironomidae & \textit{ Eukiefferiella claripennis }\\
BOLD:AAA7470 & Insecta & Diptera & Calliphoridae & \textit{ Lucilia }\\
BOLD:ACJ3716 & Insecta & Hymenoptera & Ceraphronidae & \textit{ None }\\
\hline
\end{tabular}
\label{tab:cosmos}
\end{table}

All cosmopolitans belong to different genera with the exception of two Chironomids (\textit{Limnophyes}), though 53\% of all genera present in the data are represented by a single species. Chironomidae and Cecidomyiidae are the only two families with more than two cosmopolitan members, and these are the two most diverse families sampled. 

\subsection{Environmental Indicators}
Our proposed approach co-clusters samples and species in terms of common factors. Species that share factor patterns form subcommunities, and samples that share factor patterns describe the implicit niche of each subcommunity. Although precise characterization of the implicit niches is not possible, we can study the geographic distributions of sample factors to identify regions of common profile and relate the sample factors to observable habitat covariates that can signal when a particular niche or subcommunity is likely present. 

\begin{figure}[h]
\includegraphics[width=1\linewidth]{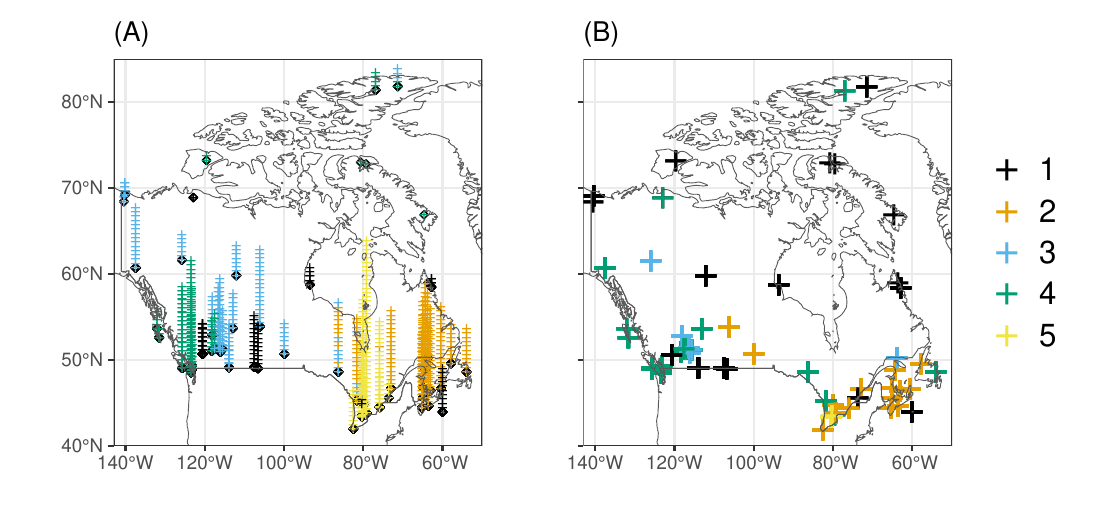}
\caption{Sample clusters in the Canada GMTP data based on estimated sample factors (A, left) and based on habitat alone (right) ($k=5$). Two points of discrepancy to attend to are 1) samples from the same site that differ in cluster membership and 2) mismatches between expected cluster assignment (B) and realized cluster assignment (A). }
\label{fig:paired_maps_k5}
\end{figure}

Sample factors $\Omega$ ``score'' each sample in terms of the various factors. The sparse estimate $\hat{O}$ directly clusters samples. Figure \ref{fig:paired_maps_k5} (A) presents a map of samples colored by the dominant factor, from which regions of common profile can be read off. Three of the sample clusters (2, 4, and 5) form distinct eco-regions organized longitudinally. Factor 2 predominates in eastern coastal Canada, factor 4 predominates in western coastal Canada, and factor 5 clusters around the Great Lakes region. Factors 1 and 3 are less geographically clustered. However, all factors tend to cluster within sites; the dominant factor is usually consistent across samples within each site, as evidenced by the consistent cluster labels. This suggests that sample factors represent geographic gradients moreso than temporal gradients.  

\begin{figure}[h]
\centering
\includegraphics[width=.65\linewidth]{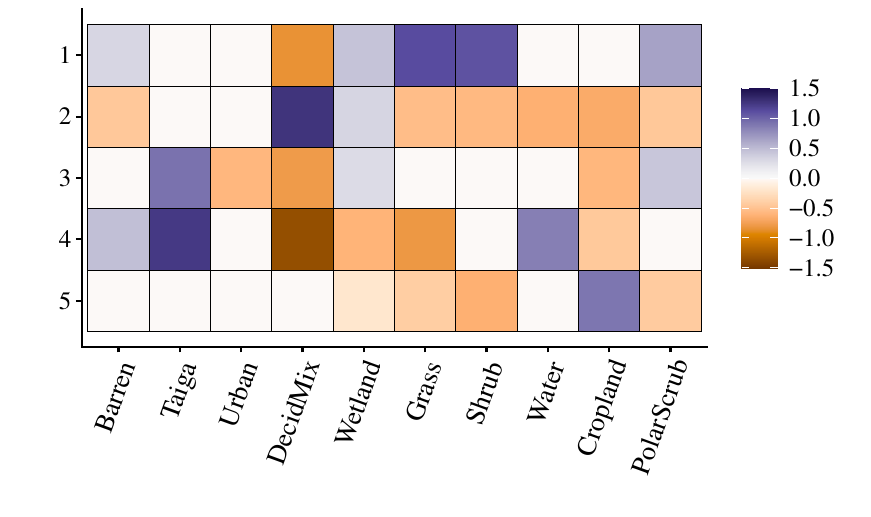}
\caption{Latent regression coefficients for each factor in our analysis of the Canada GMTP data ($k=5$). Tiles for statistically supported covariate effects are colored. }
\label{fig:coefficients_k5}
\end{figure}

Figure \ref{fig:coefficients_k5} displays estimates of statistically-supported habitat covariate effects on the sample factors. From this, we can read off the habitat types that indicate each factor. The habitat indicators of each factor are
\begin{itemize}
    \item[1. ] Open vegetation, especially grass and shrubs. See also factor five, which contrasts this factor. 
    \item[2. ] Eastern mixed deciduous forest. For no other factor is deciduous forest a positive indicator.
    \item[3. ] Inland and Polar coniferous forest.
    \item[4. ] Coastal and low-lying coniferous forest.
    \item[5. ] Cropland. Notably, the effects of other nominally similar habitats (grass, shrub, wetland, and scrub) are negative, suggesting that a specific, possibly anthropogenic feature of agriculture plays an important role.
\end{itemize}
These indicators are jointly informed by covariate effects and geographic distribution. Because covariates enter the prior for but do not strictly constrain $\Omega$, there may be a mismatch between the estimated factor and the expected factor based on covariates alone. The expected dominant factor given covariates alone is displayed in Figure \ref{fig:paired_maps_k5} (B). Habitat covariates are constant across samples from each site. There is broad agreement between (A) and (B), with notable differences between the two in Polar sites. Additionally, factor 5 (cropland, yellow) is expected to dominate in fewer sites than estimated. Nonetheless, habitat covariates are very informative about sample factors. 

\subsection{Subcommunity indicators} 
For each factor, Table \ref{tab:allIndicators_k5} lists the five highest value indicators. In doing so, we note that only nine of these 25 indicators possess Latin species names. The other 16 species may come with high indicator value but still lack species descriptions, rendering them impossible to identify for even a skilled taxonomist. Twenty of the indicators belong to the order Diptera, and of these, 13 are Chironomids. Based on their taxonomy, all of these species are small and nondescript. In practice, they will be hard to identify by anyone but the most highly skilled taxonomist,  specialized  on the family or even genus in question. This level of time and resource commitment is comparable to that of resolving and screening the full community by high-throughout, DNA-based methods and is therefore self-defeating.

Table \ref{tab:namedIndicators_k5} lists top indicators found after filtering out unnamed species. A documented species name signals that a species has been previously studied. Because larger and more distinctive species are more studied generally, named species are often larger and more distinctive than unnamed species. Hence, the presence of a taxonomic name proxies other desirable properties of a useful indicator. Two examples here are the willow leafblotch miner moth (\textit{Micrurapteryx salicifoliella}) and twenty-spotted lady beetle (\textit{Psyllobora vigintimaculata}). The former is the sixth-best indicator for factor 3 and sports a relatively large wingspan. The lady beetle, which indicates factor 4, has distinctive markings that would aid field identification. However, many other named indicators are not so charismatic, including several midges and fungus gnats that may be impossible to identify by eye. 

\begin{table}[h]
\centering
\caption{($k=5$) Top five indicators for each factor/subcommunity. ``Rank'' indicates the placement of the named indicator among all candidates. Because all species are candidates, ranks are all five or less.}
\label{tab:allIndicators_k5}
\resizebox{\columnwidth}{!}{
\footnotesize\begin{tabular}{lllllll}
\hline
\textbf{Factor} & \textbf{BIN} & \textbf{Class} & \textbf{Order} & \textbf{Family} & \textbf{Genus species} & \textbf{Rank} \\
\hline
\multirow{5}{*}{1} & BOLD:AAP6583 & Insecta & Diptera & Chironomidae & \textit{Parakiefferiella scandica} & 1\\
 &  BOLD:AAB0377  &  Insecta  &  Diptera  &  Chironomidae  &  \textit{ Smittia sp. ES12 } &  2 \\
 &  BOLD:AAA5299  &  Insecta  &  Diptera  &  Chironomidae  &  \textit{ Cricotopus } &  3 \\
 &  BOLD:AAD7251  &  Insecta  &  Diptera  &  Chironomidae  &  \textit{ Procladius dentus } &  4 \\
 &  BOLD:AAN5526  &  Insecta  &  Diptera  &  Dolichopodidae  &  \textit{ None } &  5 \\
\hline
\multirow{5}{*}{2} & BOLD:AAN5392 & Insecta & Diptera & Chironomidae & \textit{Limnophyes} & 1\\
 &  BOLD:ACK8120  &  Insecta  &  Hymenoptera  &  Platygastridae  &  \textit{ None } &  2 \\
 &  BOLD:AEV9755  &  Insecta  &  Diptera  &  Chironomidae  &  \textit{ Gymnometriocnemus brumalis } &  3 \\
 &  BOLD:ADE2870  &  Insecta  &  Diptera  &  Cecidomyiidae  &  \textit{ None } &  4 \\
 &  BOLD:AEW5280  &  Insecta  &  Diptera  &  Chironomidae  &  \textit{ Gymnometriocnemus } &  5 \\
\hline
\multirow{5}{*}{3} & BOLD:ACG1817 & Insecta & Diptera & Chironomidae & \textit{Heterotrissocladius oliveri} & 1\\
 &  BOLD:ACT6261  &  Insecta  &  Diptera  &  Chironomidae  &  \textit{ None } &  2 \\
 &  BOLD:AAP9896  &  Insecta  &  Diptera  &  Sciaridae  &  \textit{ None } &  3 \\
 &  BOLD:AAB4394  &  Insecta  &  Hemiptera  &  Aphididae  &  \textit{ Chaitophorus neglectus } &  4 \\
 &  BOLD:AAG1768  &  Insecta  &  Diptera  &  Muscidae  &  \textit{ Hydrotaea scambus } &  5 \\
\hline
\multirow{5}{*}{4} & BOLD:ACT3603 & Insecta & Diptera & Chironomidae & \textit{None} & 1\\
 &  BOLD:ACC8307  &  Insecta  &  Diptera  &  Chironomidae  &  \textit{ Gymnometriocnemus } &  2 \\
 &  BOLD:ACX1465  &  Insecta  &  Psocodea  &  Trogiidae  &  \textit{ Cerobasis guestfalica } &  3 \\
 &  BOLD:ACD9444  &  Collembola  &  Entomobryomorpha  &  Entomobryidae  &  \textit{ Entomobrya intermedia } &  4 \\
 &  BOLD:ACX6073  &  Collembola  &  Entomobryomorpha  &  Entomobryidae  &  \textit{ None } &  5 \\
\hline
\multirow{5}{*}{5} & BOLD:ABU5526 & Insecta & Diptera & Chironomidae & \textit{None} & 1\\
 &  BOLD:ACA7493  &  Insecta  &  Diptera  &  Chironomidae  &  \textit{ None } &  2 \\
 &  BOLD:ABU5520  &  Insecta  &  Diptera  &  Sciaridae  &  \textit{ Corynoptera } &  3 \\
 &  BOLD:ABU5521  &  Insecta  &  Diptera  &  Sciaridae  &  \textit{ Corynoptera furcata } &  4 \\
 &  BOLD:AAV1136  &  Insecta  &  Diptera  &  Scatopsidae  &  \textit{ Swammerdamella } &  5 \\
\hline
\end{tabular}}
\label{tab:tax}
\end{table}

\begin{table}[h]
\centering
\caption{($k=5$) Top five named indicators for each factor/subcommunity. ``Rank'' indicates the placement of the named indicator among all candidates. Because candidate species are filtered to include only named species, some ranks are greater than five. }
\label{tab:namedIndicators_k5}
\resizebox{\columnwidth}{!}{
\footnotesize\begin{tabular}{lllllll}
\hline
\textbf{Factor} & \textbf{BIN} & \textbf{Class} & \textbf{Order} & \textbf{Family} & \textbf{Genus species} & \textbf{Rank} \\
\hline
\multirow{5}{*}{1} & BOLD:AAP6583 & Insecta & Diptera & Chironomidae & \textit{Parakiefferiella scandica} & 1\\
 &  BOLD:AAD7251  &  Insecta  &  Diptera  &  Chironomidae  &  \textit{ Procladius dentus } &  4 \\
 &  BOLD:AAC3084  &  Insecta  &  Diptera  &  Chironomidae  &  \textit{ Cryptotendipes darbyi } &  15 \\
 &  BOLD:AAW3972  &  Insecta  &  Diptera  &  Chironomidae  &  \textit{ Chironomus athalassicus } &  18 \\
 &  BOLD:AAG8587  &  Insecta  &  Lepidoptera  &  Blastobasidae  &  \textit{ Pigritia murtfeldtella } &  19 \\
\hline
\multirow{5}{*}{2} & BOLD:AEV9755 & Insecta & Diptera & Chironomidae & \textit{Gymnometriocnemus brumalis} & 3\\
 &  BOLD:ACC7426  &  Insecta  &  Coleoptera  &  Cantharidae  &  \textit{ Malthodes fragilis } &  6 \\
 &  BOLD:ACA3052  &  Insecta  &  Coleoptera  &  Curculionidae  &  \textit{ Isochnus sequensi } &  11 \\
 &  BOLD:AAA9270  &  Insecta  &  Lepidoptera  &  Crambidae  &  \textit{ Scoparia penumbralis } &  14 \\
 &  BOLD:AAH3983  &  Insecta  &  Diptera  &  Sciaridae  &  \textit{ Ctenosciara hyalipennis } &  15 \\
\hline
\multirow{5}{*}{3} & BOLD:ACG1817 & Insecta & Diptera & Chironomidae & \textit{Heterotrissocladius oliveri} & 1\\
 &  BOLD:AAB4394  &  Insecta  &  Hemiptera  &  Aphididae  &  \textit{ Chaitophorus neglectus } &  4 \\
 &  BOLD:AAG1768  &  Insecta  &  Diptera  &  Muscidae  &  \textit{ Hydrotaea scambus } &  5 \\
 &  BOLD:AAD5801  &  Insecta  &  Lepidoptera  &  Gracillariidae  &  \textit{ Micrurapteryx salicifoliella } &  6 \\
 &  BOLD:AAP8779  &  Insecta  &  Diptera  &  Sciaridae  &  \textit{ Camptochaeta delicata } &  10 \\
\hline
\multirow{5}{*}{4} & BOLD:ACX1465 & Insecta & Psocodea & Trogiidae & \textit{Cerobasis guestfalica} & 3\\
 &  BOLD:ACD9444  &  Collembola  &  Entomobryomorpha  &  Entomobryidae  &  \textit{ Entomobrya intermedia } &  4 \\
 &  BOLD:AAM0871  &  Insecta  &  Diptera  &  Chironomidae  &  \textit{ Hydrobaenus fusistylus } &  6 \\
 &  BOLD:ACX4194  &  Insecta  &  Diptera  &  Sciaridae  &  \textit{ Hyperlasion wasmanni } &  8 \\
 &  BOLD:AAU2688  &  Insecta  &  Coleoptera  &  Coccinellidae  &  \textit{ Psyllobora vigintimaculata } &  9 \\
\hline
\multirow{5}{*}{5} & BOLD:ABU5521 & Insecta & Diptera & Sciaridae & \textit{Corynoptera furcata} & 4\\
 &  BOLD:AAN6447  &  Insecta  &  Diptera  &  Sciaridae  &  \textit{ Corynoptera perpusilla } &  8 \\
 &  BOLD:ABW1379  &  Insecta  &  Diptera  &  Chloropidae  &  \textit{ Malloewia abdominalis } &  9 \\
 &  BOLD:ABU5533  &  Insecta  &  Diptera  &  Phoridae  &  \textit{ Megaselia aristalis } &  12 \\
 &  BOLD:AAN6435  &  Insecta  &  Diptera  &  Sciaridae  &  \textit{ Bradysia angustipennis } &  14 \\
\hline
\end{tabular}}
\end{table}

\subsubsection{Evaluating indicators through conditional prediction}

The utility of indicator species also depends on their ability to predict distributions of unobserved species in a sample. Here, we ask: \textit{Does observing indicators aid prediction?} and \textit{Are some species harder to indicate/predict than others?} The questions are approached through the lens of conditional prediction: given observations of a subset of the community (here, indicator species), what other community members are likely also present? It is also natural to consider the extent to which indicator species improve predictions compared to using only sample indicators, like habitat covariates. 

Our predictions therefore take two forms. Given community data $(Y,X)$ and the corresponding posterior $p\left(\Gamma, \Omega, B\mid Y, X\right)$, the predictive distribution of a new sample $\bm{y}_*$ is conditioned either on only habitat information about the new sample $\bm{x}_*$, or both habitat information and indicator species data $\bm{y}_{*\mathcal{J}}$. Both distributions rely on $\bm\omega_*$. Without indicator species data, the predictive distribution of new sample factors $p(\bm\omega_*\mid \bm{x}_*, \Gamma, \Omega, B, Y)=p(\bm\omega_*\mid \bm{x}_*, B)$ can be computed directly. Given partial data, we compute the density $p(\bm\omega_*\mid \bm{x}_*,\bm{y}_{*\mathcal{J}}, \Gamma, \Omega, B, Y)=p(\bm\omega_*\mid \bm{x}_*,\bm{y}_{*\mathcal{J}}, \Gamma, B)$ using Gibbs sampling, alternating between sampling $\{\{y_{*jl}\}_{l=1}^k\}_{j\in\mathcal{J}}$ and $\bm\omega_*$ conditioned on $\bm{y}_{*\mathcal{J}}$, $\Gamma$, and $B$. Given $\bm\omega_*$, $\bm{y}_*$ follows the usual Poisson. To imitate out-of-sample data, we perform 10-fold cross validation. 

We use three strategies for choosing $\mathcal{J}$ based on MB-IndVal. The first considers all species and selects the 15 highest-scoring species for each factor $\mathcal{J}^{(1)}$. As noted above, however, some species are not practical indicators due to their small size or novelty. The second and third strategies restrict the set of candidate indicators to charismatic taxa and physically large specimens. ``Charismatic'' indicators $\mathcal{J}^{(2)}$ are the highest scoring named species belonging to Lepidoptera (moths and butterflies) or Coleoptera (beetles). ``Large'' indicators $\mathcal{J}^{(3)}$ are the highest scoring named species in the 90th body-size percentile. Body sizes are obtained from BIOSCAN-5M \citep{gharaee2024bioscan}, which includes standardized images of a large collection of arthropods. Not all species studied here are included in BIOSCAN-5M, which further restricts the set of large candidates. 

\begin{figure}[h]
    \centering
    \includegraphics[width=1.0\linewidth]{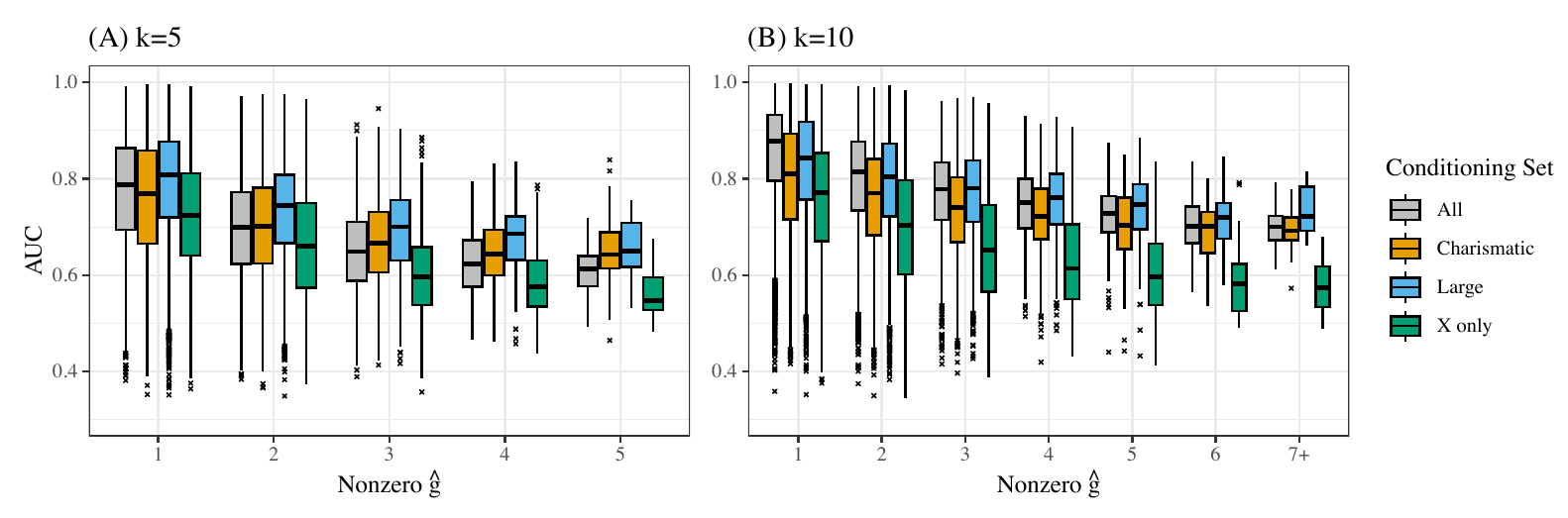}
    \caption{Results for predicting the non-indicator species based on the habitat of the sample ($X$) and the indicator species, for different restrictions on the candidate indicator species (conditioning set).}
    \label{fig:AUC_k5and10}
\end{figure}

In making comparisons, we consider $k=5$, as used in other analyses, and $k=10$, as is preferred by WAIC. The chosen prediction metric is AUC, which respects the data set's extreme sparsity and the primary interest in predicting other species' occurrences rather than trap counts. For each species, we compute the average AUC across held-out samples and stratify the species-specific AUCs by the number of factors that each species loads on, i.e., $||\hat{\bm{g}}_j||_0$. This stratification is interesting because MB-IndVal is factor-specific and favors species that load on few factors (specificity). It would therefore be natural for indicators chosen using MB-IndVal to predict non-specialists less well than specialists. Figure \ref{fig:AUC_k5and10} displays the results. Three results are apparent: 1) all predictions worsen as $||\hat{\bm{g}}_j||_0$ grows, 2) conditioning on 15 indicators per factor improves on predictions that condition only on covariates, and the degree of improvement grows slightly with $||\hat{\bm{g}}_j||_0$, and 3) the most predictive choice of $\mathcal{J}$ depends on $||\hat{\bm{g}}_j||_0$. Result (1) is evidenced by the fairly consistent downward trend in AUC across conditioning sets and $k$. Species that load on many factors have more complex distributions and are more difficult to predict. Result (2) is not unexpected, but covariate-only predictions are impressively accurate for $\{j: ||\hat{\bm{g}}_j||_0=1\}$, which agrees with the previous observation that $X$ is fairly predictive of $\Omega$. Lastly, (3) is somewhat surprising in that charismatic and large species are often more predictive than $\mathcal{J}^{(1)}$. Predictions conditioned on $\mathcal{J}^{(1)}$ compare most favorably to $\mathcal{J}^{(2)}$ and $\mathcal{J}^{(3)}$ predictions for small $ ||\hat{\bm{g}}_j||_0$. This too is consistent with the construction of MB-IndVal, which prizes species that are specific to individual factors. By restricting the set of candidate indicators, species that are less factor-specific are chosen as indicators, which aids predictions of other factor-nonspecific species. Hence, although MB-IndVal is a valuable tool for indicating specialist species, another metric may be more useful for other species.  

The difference between covariate-only predictions and indicator species predictions is also moderated by the number of indicator species used. While 15 indicators per factor represents a nearly 1000-fold reduction in the number of species to monitor, resolving $\sim100$ arthropod species is still a significant challenge. With $k=10$ and 15 indicators ($\mathcal{J}^{(1)}$), the average AUC across all samples and species 0.814, and the average covariate-only AUC is 0.716, a marked decrease. Using only five indicators, however, the indicator-based AUC drops to 0.729, with the difference naturally more pronounced for some community subsets than others.

\section{Discussion}
Modern biodiversity data and the many unseen species therein present an opportunity to better understand prevailing species subcommunities and ecosystem types across the globe. Yet, the ecological and statistical complexity inherent to such large-scale data present serious obstacles. \textit{Ad hoc} species or site classifications may be uninformative, especially in previously un- or under-studied systems. Statistical or model-based methods for this kind of nuanced clustering are also underdeveloped. 

Our approach to learning dominant ecosystem and species types is to first construct a flexible and scalable model for large community data. Poisson factorization can be made very flexible through the use of hierarchical priors and can handle very large, sparse data. It also has the important added benefit of scoring samples and species in terms of a small number of shared additive latent factors. The prior introduces covariates as candidate environmental indicators for each factor, and Bayesian decoupling maps sample and species scores to precise clusters, thereby identifying dominant subcommunities and environment types. Lastly, we derive a data-driven ranking system for indicator species, adapting a classical framework to modern data settings. 

Application of this approach led to partial answers to our initial research questions:
\begin{itemize}
   \item[1) ] \textbf{How can Canadian arthropods be partitioned into ecologically-relevant subcommunities?} Using sparse species factors, we find that approximately half of the community can be partitioned into five distinct subcommunities. The remaining species belong to or overlap with two or more of the subcommunities. 
   \item[2) ] \textbf{To what extent does this partition reflect observed environmental gradients? Can subcommunity presence or species occurrence be predicted using environmental indicators?} The five subcommunities are aligned with habitats indicated by open vegetation, mixed deciduous forests, inland and polar conifer forests, coastal and low-lying conifer forests, and crop land. These environmental indicators are generally predictive of site clusters and are most predictive of species occurrence for species belonging to the distinct subcommunities. Sample clusters also exhibit varying degrees of geographic clustering---three are tightly clustered and organized longitudinally, whereas two others are more geographically heterogeneous. 
   \item[3) ] \textbf{What species indicate each subcommunity? Can subcommunity presence or species occurrence be predicted using indicator species?} We identify indicator species for each subcommunity using MB-IndVal. However, nearly all indicator species are totally novel or too small and similar to distinguish except through DNA, a feature to be expected in most studies like GMTP. Yet, small numbers of indicator species can effectively predict occurrences of unmeasured species and can do so better than environmental indicators alone. Indicators are best suited to predicting species that occur in specific ecosystems.
\end{itemize}

Undoubtedly, selecting strong, data-driven indicators is challenging and deserves further study. However, our results suggest some general conclusions: 1) Both indicator species and environmental indicators are important, and they are best used in tandem---in our analysis, habitat alone was an effective predictor of ecosystem type and species occurrence, and indicator species improve these predictions, provided sufficiently many indicators are considered; 2) Even when using all available indicators, some species will be poorly indicated---model-based approaches for selecting indicators enable the user to identify these blind spots; 3) Diverse indicators are needed to monitor diverse communities and ecosystems---simply selecting a set of ``good'' indicators for each subcommunity and habitat type, like we do here, overlooks species with more nuanced environmental preferences and responses. With these conclusions and the present toolbox in mind, we propose using 10–15 large indicators for each subcommunity and habitat information to represent and predict some Canadian arthropods, namely those 6172 species within the sharp subcommunities. If none of the 10–15 large indicators are found, then environmental indicators will suffice. 

A model that preserves the features described here, adapted to other types of biodiversity data, like presence-absence, would be valuable. This could be achieved by modeling occurrence $z_{ij}$ as $z_{ij}=\mathbbm{1}(y_{ij}>0)$ and modeling $y_{ij}$ as described in this paper. The MB-IndVal score we describe can also be extended in a number of ways. Most pressingly needed is a way to jointly score sets of indicator species that avoids selecting redundant indicators, possibly building on \cite{de2010improving}. Relatedly, directly incorporating the practicality of using a species as an indicator (e.g., size, distinctiveness) into its indicator value will be necessary to avoid complications presented by small and novel species. 

\section*{Acknowledgments}
This research was partially supported by the European Research Council under the European Union’s Horizon 2020 research and innovation programme (grant agreement No 856506) and the National Science Foundation (IIS-2426762).

\section*{Data availability}
Data are available as datasets DS-20GMP01 through to DS-20GMP37 on BOLD, which can be downloaded by dataset code via the \verb|BOLDconnectR| package \citep{boldconnect2025}. For protocol details, visit \url{https://biodiversitygenomics.net/projects/gmp/}. See also \cite{seymour2024global}.

\newpage

\bibliography{bibliography}       

\newpage


\section*{Convergence assessments}
\begin{figure}[h]
    \centering
    \includegraphics[width=0.65\linewidth]{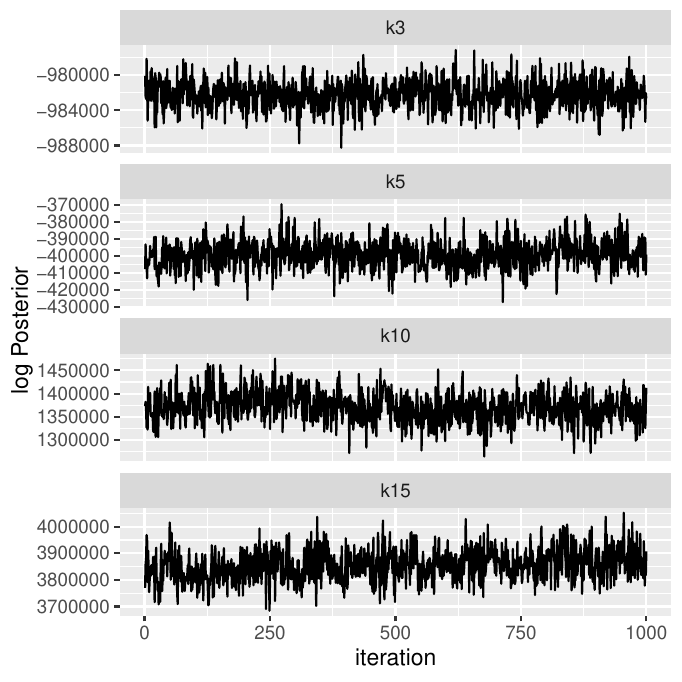}
    \caption{Traceplots of log posterior density for different model ranks. Effective sample sizes (ESS) are 1000,  626,  169, and 238. }
    \label{fig:supp_logPostTrace}
\end{figure}

\begin{figure}[h]
    \centering
    \includegraphics[width=0.85\linewidth]{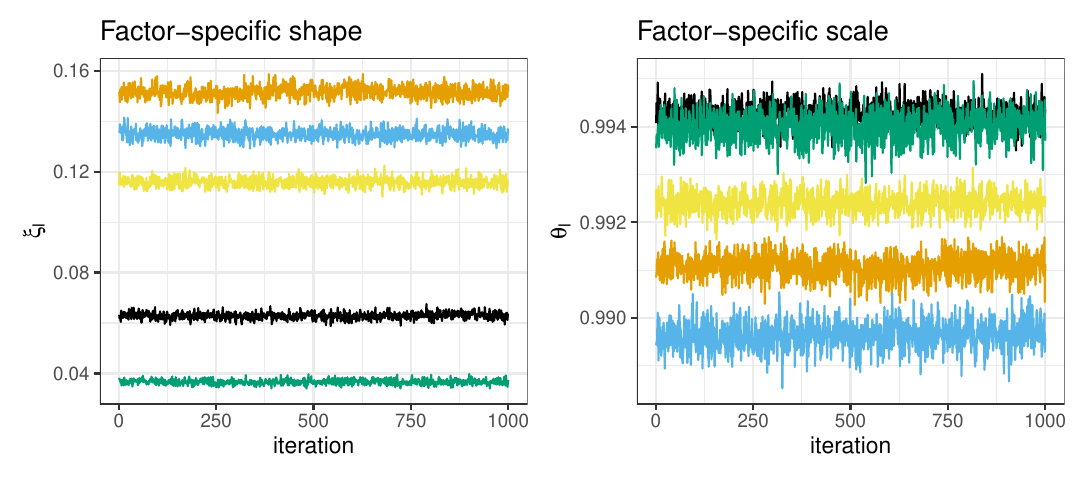}
    \caption{Trace plots for shape and scale parameters of species factor loadings prior distributions ($k=5$). Although some scale traces overlap, shape traces suggest no label switching occurs. }
    \label{fig:supp_hyperparameterTrace}
\end{figure}

\begin{figure}[h]
    \centering
    \includegraphics[width=0.85\linewidth]{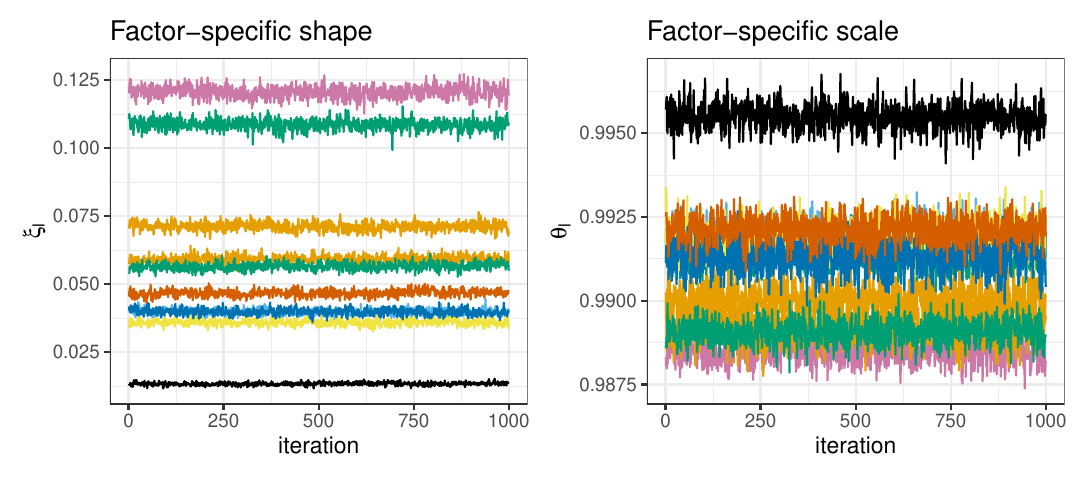}
    \caption{Trace plots for shape and scale parameters of species factor loadings prior distributions ($k=10$). Although some shape and scale traces overlap, shape and scale traces do not simultaneously overlap, indicating no label switching occurs.}
    \label{fig:supp_hyperparameterTrace_k10}
\end{figure}

We provide trace plots of the log joint posterior density for $k\in\{3,5,10,15\}$ (figure \ref{fig:supp_logPostTrace}), and trace plots of species factor hyperparameters $\bm\xi$ and $\bm\theta$ for $k=5$ and $k=10$ (figures \ref{fig:supp_hyperparameterTrace} and \ref{fig:supp_hyperparameterTrace_k10}. Effective log posterior sample sizes are 1000, 626, 169, and 238, respectively. We only construct estimates that are sensitive to factor permutation (label switching) using models with $k=5$ and $k=10$. The hyperparameter traceplots indicate no signs of permutation during sampling.

\newpage

\section*{Model assessment and comparison}
\begin{figure}[h]
\centering
\includegraphics[width=.65\linewidth]{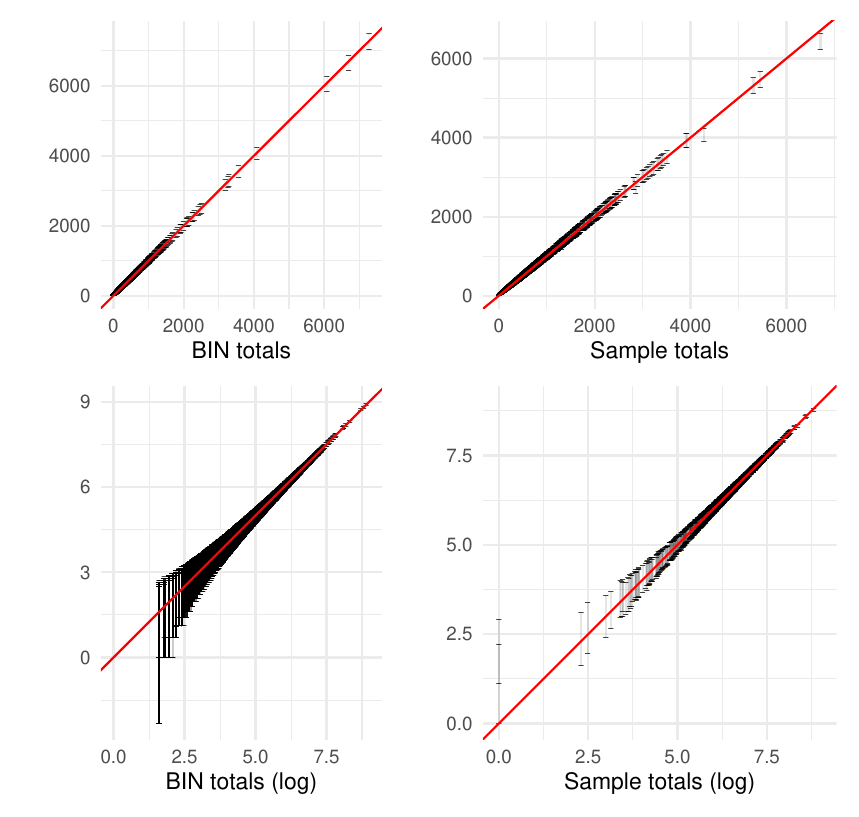}
\caption{Posterior predictive row (sample) and column (species/BIN) marginals versus empirical marginals for $k=5$ on the original scale (top) and log scale (bottom).}
\label{fig:supp_postpred_k5}
\end{figure}

Four model ranks are considered and compared with WAIC (table \ref{tab:ICs}. Rank five is preferred to ranks three or 15, and rank 10 is preferred to five. Figure \ref{fig:supp_postpred_k5} displays posterior predictive row and column marginal distributions against observed marginals on both original and log scales. The posterior predictive intervals contain the true marginals almost unanimously. 

\begin{table}[h]
  \centering
  \caption{WAIC comparison for five different model ranks.}
  \label{tab:ICs}
  \begin{tabular}{c|c c}
    \hline
    \textbf{k} & \textbf{WAIC}  \\
    \hline
    $k=3$  & 2,332,045 \\
    $k=5$  & 1,909,514 \\
    $k=10$ & \textbf{1,445,511} \\
    $k=15$ & 2,135,923 \\
    \hline
  \end{tabular}
\end{table}

\end{document}